\documentclass[11pt,a4paper]{article}
\pdfoutput=1
\usepackage{jcappub}
\usepackage{slashed}
\usepackage{stackengine}
\def\be{\begin{equation}}
\def\ee{\end{equation}}
\def\bea{\begin{eqnarray}}
\def\eea{\end{eqnarray}}

\begin{document}

\title{Information Graph Flow: a geometric approximation of quantum and statistical systems} 

\author{Vitaly Vanchurin}

\emailAdd{vvanchur@d.umn.edu}

\date{\today}

\affiliation{Department of Physics, University of Minnesota, Duluth, Minnesota, 55812 \\
Duluth Institute for Advanced Study, Duluth, Minnesota, 55804}

\abstract{Given a quantum (or statistical) system with a very large number of degrees of freedom and a preferred tensor product factorization of the Hilbert space (or of a space of distributions) we describe how it can be approximated with a very low-dimensional field theory with geometric degrees of freedom. The geometric approximation procedure consists of three steps. The first step is to construct weighted graphs (we call information graphs) with vertices representing subsystems (e.g. qubits or random variables) and edges representing mutual information (or the flow of information) between subsystems. The second step is to deform the adjacency matrices of the information graphs to that of a (locally) low-dimensional lattice using the graph flow equations introduced in the paper. (Note that the graph flow produces very sparse adjacency matrices and thus might also be used, for example, in machine learning or network science where the task of graph sparsification is of a central importance.) The third step is to define an emergent metric and to derive an effective description of the metric and possibly other degrees of freedom. To illustrate the procedure we analyze (numerically and analytically) two information graph flows with geometric attractors (towards locally one- and two-dimensional lattices) and metric perturbations obeying a geometric flow equation. Our analysis also suggests a possible approach to (a non-perturbative) quantum gravity in which the geometry (a secondary object) emerges directly from a quantum state (a primary object) due to the flow of the information graphs.}

\maketitle

\section{Introduction}
Quantum gravity is a hypothesis that the gravitational degrees of freedom can somehow be described by vectors in Hilbert space $|\psi\rangle$ and that a unitary evolution of these vectors in certain limits can be approximated by the Einstein field equations. Exactly how this should be carried out in practice is unclear, but at least in context of the AdS/CFT \cite{Maldacena, Witten} some rules were spelled out \cite{Banks, Harlow}. The main idea is that the Hilbert space and unitary evolution is that of a conformal field theory (CFT) without gravity and the gravitational and other degrees of freedom on the Anti-de-Sitter (AdS) space are specified by a mapping of observables from CFT to AdS (also known as the AdS/CFT dictionary). If the mapping is in fact one-to-one then there is really only a single quantum theory, but there are two ways to look at its predictions and observables, i.e. the CFT way and the AdS way. In this paper we will argue that there can be many more (low-dimensional field theory) effective descriptions of the same quantum system defined by different graph flow equations, but since these descriptions are only approximate the mappings between the field theories may or may not be one-to-one. 

An important point to make is that the unitary evolution of the CFT introduces a preferred tensor product factorization of the Hilbert space into ``local'' degrees of freedom and thus gives rise to a preferred set of basis vectors and consequently defines a wave-function $\psi(\vec{x}) =  \langle \vec{x} |\psi\rangle$ with normalization $\int d^Nx\; \psi^*(\vec{x}) \psi(\vec{x}) =1$. (Note that in the case when components of the vector $\vec{x}$ take only discrete set of values, as in the case of qubits, the normalization integral should be interpreted as a sum.) In what follows we will also assume that such a factorization of the Hilbert exists, but we will make no assumption of the locality, i.e. $x_a$ need not to have any special relation to $x_{a+1}$). Thus our starting point is not CFT, but an arbitrary quantum state and a set of orthonormal basis vectors which when combined define a quantum wave-function $\psi(\vec{x})$.  An important question then one might ask is how to assign a semi-classical geometry to a given wave-function $\psi(\vec{x})$? In fact the same question can be asked about an arbitrary probability distribution $p(\vec{x})$ with the only difference that it is real and normalized as  $\int d^Nx\; p(\vec{x})  =1$. 

In the recent years a number of interesting proposal was put forward to address this issue. One set of ideas is based on a special kinds of quantum circuits (known as tensor networks) which were first used to study quantum many-body systems  \cite{Vidal}, but eventually found their way into quantum gravity research \cite{Swingle}. Indeed, quantum circuits provide an interesting new prospective on how the geometry (e.g. AdS) might emerge from a given quantum state (e.g. CFT) by identifying classical space with a tensor network which would produce such state. It was also realized that holographic CFT states can be represented in terms of error-correcting codes which suggests a possible mechanism for the emergence of bulk locality  \cite{Dong} and an approximate \cite{Harlow2} Ryu-Takayanagi formula \cite{Ryu}. In Refs  \cite{Susskind} the authors went further and conjectured that the complexity of holographic CFT states is dual to a field theory actions over certain patches in the AdS spacetime. In Ref. \cite{Vanchurin} we extended their analysis by defining a high-dimensional dual field theory whose Euclidean action is proportional to computational complexity of a (shortest) quantum circuit which takes an arbitrary initial state (not necessarily product state) to an arbitrary final state (not necessarily holographic state). It would be interesting to see if the high-dimensional dual field theory without gravity introduced in Ref. \cite{Vanchurin} could be approximated as a low-dimensional theory with gravity using the geometric approximation procedure introduced in this paper.

Another approach to quantum gravity is based on a direct analysis of the entanglement structure of quantum states. For example, in Ref. \cite{Carroll} the authors used the procedure of multidimensional scaling to generate the best-fit spacial geometry from a weighted graph defined by the metric of mutual information between subsystems; and in Ref. \cite{Noorbala} the author described a procedure of generating a family of spacial geometries from different tensor product decompositions of a Hilbert space which suggests a possible mechanism for the emergence of time. In this paper we will also use the entanglement structure to form weighted graphs (we call information graphs), but we will not attempt to embed the graph in a higher-dimensional space and we will leave the tensor product decomposition of the Hilbert space fixed. Instead, we will continuously deform the adjacency matrices of the information graphs to that of a (locally) low-dimensional lattice using the graph flow equations. Although the initial discussion will be centered around quantum systems of qubits or a statistical system of bits, the procedure can be applied to arbitrary wave-function $\psi(\vec{x})$ or probability distributions $p(\vec{x})$. In either case the mutual information (or the flow of information) between subsystems would be used to infer the best-fit low-dimensional field theory with geometric degrees of freedom.

On the following chart we summarize various approximations steps  (horizontal direction) that we shall take as well as the temporal dynamics (vertical direction) of either quantum or statistical system that we are trying to model
\\
\bea
|\psi\rangle(0)\, \text{or} \, \vec{p}(0)   & \xrightarrow{\text{information graph, Sec \ref{Sec:Information Graph}}}  \hat{A}(0,0)     \xrightarrow{\text{graph flow, Sec \ref{Sec:GraphFlow}}}  \hat{A}(0,\tau)   \xrightarrow{\text{emergent geometry, Sec \ref{Sec:EmergentGeometry}}}  g_{IJ}(0) & \notag\\
\Big{\downarrow}\; \text{\scriptsize \stackanchor{unitary}{dynamics}}\; \Big{\downarrow} &  \;\;\;\;  \;\;\;\;\;\; \;\;\;\;\;\; \;\;\;\;\;\; \;\;\;\;  \;\;\;\;  \Big{\downarrow}\; \text{\scriptsize \stackanchor{information}{dynamics }}\; \;   \;\;\;\;\;\; \;\;\;\;\;\; \;\;\;\;\;\;  \Big{\downarrow}\;\text{\scriptsize \stackanchor{approximate}{dynamics } }  \;\;\;\;\;\; \;\;\;\;\;\;\text{\scriptsize \stackanchor{emergent}{dynamics} }  \Big{\downarrow} & \;\;\;\; \;  \notag\\
|\psi\rangle(t)\,\text{or} \,\vec{p}(t)   &  \xrightarrow{\text{information graph, Sec \ref{Sec:Information Graph}}}  \hat{A}(t,0)     \xrightarrow{\text{graph flow, Sec \ref{Sec:GraphFlow}}}  \hat{A}(t,\tau)   \xrightarrow{\text{emergent geometry, Sec \ref{Sec:EmergentGeometry}}}  g_{IJ}(t) &\notag\\\notag \\
\label{eq:Chart}
\eea
Exact meaning of all these arrows should become a lot more clear as we introduce the corresponding maps in the paper. 

The paper is organized as follows. In the next section we define information graphs for quantum  and statistical systems and construct corresponding adjacency matrices. In Sec. \ref{Sec:GraphFlow} we present a set of conditions that any graph flow equation must satisfy and then describe two flows whose attractors are respectively one- and two-dimensional lattices. In Sec. \ref{Sec:EmergentGeometry} we construct an emergent metric and derive an effective description of metric perturbations which obey a Ricci flow equation. In Sec. \ref{Sec:Conclusion} we summarize the main results and describe a proposal for (a non-perturbative) quantum gravity. 

\section{Information Graph}\label{Sec:Information Graph}

In this section we provide an explicit construction of the information graphs from computational states, but the construction can be easily generalized to other quantum states or probability distributions. A reader who is mainly interested in the applications of the graph flow to the problem of graph sparsification can skip this section. 

\subsection{Probability distribution}

Consider a classical probability distributions over $N$ classical bits. The sample space of such systems is specified by $2^N$ elements, i.e. $\{0,...,0,0\}$, $\{0,...,0,1\}$,... $\{1,...,1,1\}$, which form computational basis for the space of distributions. In other words an arbitrary probability distribution over $N$ classical bits is a  $2^N$-real dimensional vector
\be 
\vec{p} = p^i  \vec{i}  = \sum_i p^i  \bigotimes_{\mu}   \vec{i}_\mu \label{eq:prob_vector}
\ee
where $\mu=0,...,N$,\footnote{To avoid confusions we will always use Greek letters (e.g. $\mu, \nu, ...$)  for indices which run from $1$ to $N$ and Latin letters (e.g. $i,j,...$) for indices which run from $0$ to $2^N-1$. Also, note that the Einstein summation convention over repeated indices is implied everywhere in the paper unless stated otherwise. } and $\bigotimes$ denotes a tensor product of the corresponding vectors. Thus if  $\vec{i}_\mu \in \{\vec{0}, \vec{1} \}$ is a unit vectors in two dimensional space (not to confuse with components), then $\vec{i} \in \{\vec{0}, \vec{1},...,\overrightarrow{2^N-1}\} $ is a unit vector in $2^N$ dimensional space, i.e. probability distributions space on $N$ bits. Since $\vec{p}$ describes a probability distribution we must  also impose a normalization condition
\be
\sum_i p_i =1.
\ee

Then a marginal probability distribution of $\vec{p}$ is calculated  by summing over a subsystem defined by an integer $k \in  \{0,1\}^N$ we call a mask.  The value of $\mu$'s binary digit (or $k_\mu$) is telling us whether $\mu$'s  qubit is summed over (if $k_\mu=0$) or not (if $k_\mu=1$). Then the marginal probability distribution with mask $k$ is given by
\be
\vec{p}[k]  \equiv  \sum_i p^{i}  \bigotimes_{\mu}    \vec{i}_\mu^{\phantom{1}k_\mu}   \label{eq:partial}
\ee
where in the last line we use a very weird, but convenient, notation $\vec{x}^{\phantom{1}0}=1$ and $\vec{x}^{\phantom{1}1}=\vec{x}$.  For example,
\be 
\vec{p}[0]= \sum_i p^{i}  \bigotimes_{\mu}   \vec{i}_\mu^{\phantom{1}0}   = 1
\ee
and
\be 
\vec{p}[2^N-1]=  \sum_i p^{i}  \bigotimes_{\mu}   \vec{i}_\mu^{\phantom{1}1} = \vec{p} 
\ee
as expected.

The next step is to construct a graph  (we call {\it information graph}) with vertices representing different bits and weighted edges representing mutual information shared by these bits. Then the off-diagonal elements of the adjacency matrix of the information graph are given by
\be
A_{\mu\nu} =S(\vec{p}[2^\mu])+S(\vec{p}[2^\nu])-S(\vec{p}[2^\mu+2^\nu]) \label{eq:offdiag}
\ee
where the so-called Shannon entropy is
\be
S(\vec{q})=-\sum_i q_i \log_2 q_i,
\ee
and it is assumed that  $0 \log_2 (0) =0$. For simplicity we set all of the diagonal elements to zero. Evidently $A_{\mu\nu}$ is a symmetric matrix which  encodes the amount of information one can learn about $\mu$'s bit by measuring $\nu$'s bit.

\subsection{Quantum state}

Now, consider a quantum system of $N$ qubits whose states are described by ket vectors in $2^N$-complex dimensional Hilbert space. In the computational basis the ket vectors can be expressed as
\be
| \psi \rangle = \psi^i |i\rangle = \sum_i \psi^i  \bigotimes_{\mu}   |i_\mu\rangle 
\ee
where ${i} \in \{0,1\}^N $, ${i}_\mu \in \{0, 1\}$ and $\mu \in \{1, ..., N\}$. The corresponding bra (or dual) vectors are defined as
\be
\langle \psi | = \langle i |  \psi_i  = \sum_i \psi_i  \bigotimes_{\mu}   \langle i_\mu| 
\ee
where $\psi^i = \psi_i^*$ and because of the normalization condition  
\bea
\langle i | j \rangle = \delta_{ij}\\
\langle \psi | \psi \rangle =1
\eea
and thus
\be
\psi^i \psi_i = 1.
\ee
One can also form corresponding (pure state) density matrices as
\be
\hat{\rho} =  |\psi \rangle \langle \psi | = \rho^i_{\phantom{i}j} |i\rangle \langle j | 
\ee
where 
\be
\rho^i_{\phantom{i}j} = \psi^i \psi_j.
\ee

Then a partial trace of the density matrix (a quantum analog of a marginal probability distribution) is defined by tracing over subsystem once again defined by a mask  $k \in  \{0,1\}^N$. The value of $\mu$'s binary digit (or $k_\mu$) is now telling us if $\mu$'s  qubit is traced over (if $k_\mu=0$) or not (if $k_\mu=1$) and  thus the reduced density matrix with mask $k$ is given by
\bea
\hat{\rho}[k] & \equiv & Tr_k (\hat{\rho}) = Tr_k(\sum_{i,j} \rho^i_{\phantom{i}j}  \bigotimes_{\mu}   |i_\mu\rangle  \bigotimes_{\nu}  \langle j_\nu|   ) \notag\\
& = & \sum_{i,j} \rho^{i}_{\phantom{i}j}  \bigotimes_{\mu}   |i_\mu\rangle^{k_\mu}  \bigotimes_{\nu}  \langle j_\nu|^{k_\nu}  \prod_\lambda (\delta^{j_\lambda}_{i_\lambda})^{1- k_\lambda}   \label{eq:partial}
\eea
where in the last line we use a (once again) weird, but convenient, notation $|\psi \rangle^1=|\psi\rangle$, $\langle \psi |^1 =\langle \psi |$ and $|\psi \rangle^0 =1 $, $\langle \psi |^0 =1$. Moreover the product of delta symbols raised to powers of $k_\mu$ ensures that correct indices are contracted in the partial trace. It follows that
\be 
\hat{\rho}[0]= \sum_{i,j}\rho^{i}_{\phantom{i}j}  \bigotimes_{\mu}   |i_\mu\rangle^0  \bigotimes_{\nu}  \langle j_\nu|^0  \prod_\lambda (\delta^{j_\lambda}_{i_\lambda})^{1}  = \rho^{i}_{\phantom{i}i} =  \psi^i \psi_i =1
\ee
and
\be 
\hat{\rho}[2^N-1]= \sum_{i,j}\rho^{i}_{\phantom{i}j}  \bigotimes_{\mu}   |i_\mu\rangle^1  \bigotimes_{\nu}  \langle j_\nu|^1  \prod_\lambda (\delta^{j_\lambda}_{i_\lambda})^{0}  = \rho^{i}_{\phantom{i}j} |i\rangle \langle j | = \hat{\rho} 
\ee
as expected. Then information graph is constructed by defining off-diagonal elements of an adjacency matrix as 
\be
A_{\mu\nu} =S(\hat{\rho}[2^\mu])+S(\hat{\rho}[2^\nu])-S(\hat{\rho}[2^\mu+2^\nu]) \label{eq:offdiag}
\ee
where 
\be
S(\hat{\sigma})=-Tr_0(\hat{\sigma} \log_2 \hat{\sigma})
\ee
is the von Neumann entropy.  All of the diagonal elements are once again set to zero. Just like in the case of probability distributions, the adjacency matrix $A_{\mu\nu}$ is a symmetric matrix and is telling us how much qubit $\mu$ is entangled with qubit $\nu$ using mutual information as a measure of entanglement. Note that the above density matrix $\hat{\rho}$ was that of a pure state $|\psi\rangle$, but for what follows the purity of the state is not need. In fact we could have started with an arbitrary density matrix (i.e. a positive semidefinite, Hermitian matrix with unit trace) and the construction of the information graph and adjacency matrix would be practically unchanged. 
 
So far the informational dependence in a given (quantum or statistical) system was described by only a single undirected graph, but there are situations when this would not be sufficient. In particular, if for a (quantum or statistical) system in question we know not only the current state, but also evolutions laws (described by a Hamiltonian or transition matrix) then information can flow from one subsystem to another. Such a flow of information need not be symmetric, but can still be described by a weighted and also directed information graph (or more generally a hyper-graph\footnote{Hyper-graph is a generalization of a graph in which an edge can join an arbitrary number of vertices.}  \cite{Severson}). For example, a non-symmetric adjacency matrix can be defined as
\be
A_{\mu\nu} = \left | \frac{d }{d t} S(\vec{p}[2^\mu]) -  \frac{d }{d t} S(\vec{p}[2^\nu])\right |
\ee
for classical probability distributions or
\be
A_{\mu\nu} = \left | \frac{d }{d t} S(\hat{\rho}[2^\mu]) -  \frac{d }{d t}S(\hat{\rho}[2^\nu]) \right |
\ee
for quantum states. The new information matrix describes the rate of flow of (positive) information (about the rest of the system) from bit $\nu$ to bit $\mu$. 

Another generalization is to allow subsystems to contain multiple bits that are distinguishable (e.g. labeled by some index) and thus the informational dependence between bits of the same type might have to be treated differently form the bits of different types. In terms of graphs this implies that one should allow a multiple number  of different information graphs (or a single graph with different colors of edges) described by different adjacency matrices. Since both generalization are quite natural in context of the graph flows introduced in the next section we shall allow the informational dependence to be described by an arbitrary finite collection of not necessarily symmetric adjacency matrices.

\section{Graph Flow}\label{Sec:GraphFlow}

For starters consider a single adjacency matrix $\hat{A}$ which may or may not be symmetric, but must have non-negative elements, i.e. $A_{ij}\geq0$. The adjacency matrix describes the structure of information graphs defined in Sec. \ref{Sec:Information Graph} which are weighted and (in some cases) directed. Then our next task is to continuously deform (or flow) it to another adjacency matrix which represents a $D$-dimensional lattice at the smallest scales, but can have a non-trivial geometry and also topology on larger scales. The main idea here is to define an evolution equation for the adjacency matrix (or what we call the graph flow equation) as a continuous dynamical system
\be
\frac{d}{d\tau} {\hat{A}} = \hat{f}(\hat{A}). \label{eq:GraphFlow}
\ee
Note that on the right hand side there is a collection of $N^2$ functions $f_{ij}$'s  of $N^2$ variables $A_{ij}$'s each, but these functions must satisfy the following constraints:

{\it (1) Permutation of vertices.} Since our labeling  of vertices (or qubits, or bits, or subsystems) is arbitrary, we expect the graph flow to be invariant under arbitrary permutations of indices describing vertices.

{\it (2) Non-negativity of elements.} By definition adjacency matrices have only non-negative elements and thus we should constraint the dynamics to only trajectories for which individual elements remain non-negative.

{\it (3) Finiteness of elements.} The adjacency matrix must also maintain finiteness of its elements throughout the evolution  which imposes the third necessary constraint on the graph flow equation. 

There might be other constraints (that are relevant for a particular application of the graph flow), but we would like to emphasize that these three constrains are already very restrictive and many simple choices of functions $f_{ij}$'s (e.g. $f_{ij}(\hat{A}) =f(\hat{A})$ or $f_{ij}(\hat{A})=f({A}_{ij})$) would not work. Luckily, there is still a lot of freedom in specifying non-trivial graph flows, but in this paper we shall only concentrate on the flows which can satisfy two more conditions: \

{\it (4) Attraction towards locally low-dimensional lattices (Sec. \ref{geometricattractor}).

 (5) Invariance under permutation of adjacency matrices (Sec. \ref{coloredgraphs}). }

\subsection{Geometric attractor}\label{geometricattractor}

Since our main goal in the paper is to approximate either quantum or statistical systems with low-dimensional geometries, we will be only interested in the graph flow equations with attraction towards the adjacency matrices of (locally) low-dimensional lattices. We will refer to such attractors as {\it geometric attractors} and to the corresponding graph sparsification  procedure as {\it geometric  sparsification}. To be more precise what will be attracted towards a low-dimensional lattice is an undirected adjacency matrix defined as
\be
\hat{\bf{A}}\equiv \hat{A}+\hat{A}^{\dagger},\label{eq:undirected}
\ee
where $\hat{A}^\dagger$ is a transpose of $\hat{A}$, even though the dynamics of the graph flow \eqref{eq:GraphFlow} is specified for matrix $\hat{A}$. We would also like to stress that the fact that $\hat{A}$ is not symmetric is essential for the geometric attractor presented here and so in certain cases an exactly symmetric $\hat{A}$ would have to be perturbed. 

An example of a graph flow with a $1D$ geometric attractor is given by the following equation
\be
\frac{d}{d\tau} {\hat{A}} = \alpha \hat{A} \star \left (\hat{A} -  \hat{A} \hat{A}^{\dagger} \hat{A} \right ). \label{eq:dimone}
\ee 
where by $\star$ we denote the Hadamard (or element-wise) product , i.e. $(\hat{A} \star \hat{B})_{ij} = \hat{A}_{ij} \hat{B}_{ij}$ with no summations over repeated indices. We shall now check that the first four conditions are satisfied and the system naturally flows to a one-dimensional lattice. 

The first condition, i.e. permutation invariance of vertices, follows from
\bea \notag
\hat{P} \hat{f}(\hat{P}^{\dagger} \hat{A} \hat{P}) \hat{P}^{\dagger} &= & \alpha\hat{P} \left ( \left (\hat{P}^{\dagger} \hat{A} \hat{P} \right ) \star \left (\hat{P}^{\dagger}\hat{A} \hat{P}- \hat{P}^{\dagger} \hat{A} \hat{P} (\hat{P}^{\dagger}\hat{A} \hat{P})^{\dagger} \hat{P}^{\dagger} \hat{A} \hat{P}\right ) \right  )  \hat{P}^{\dagger} \\
& = &  \alpha\hat{P}  \left ( \left (\hat{P}^{\dagger} \hat{A} \hat{P} \right ) \star \left (\hat{P}^{\dagger} \left (\hat{A} -  \hat{A} \hat{A}^{\dagger} \hat{A}\right ) \hat{P} \right) \right  )  \hat{P}^{\dagger}  =  \hat{f}(\hat{A}).
\eea
where $\hat{P}$ is a permutation matrix and the last step is due to invariance of the Hadamard product under actions of permutation matrices (switching the labels of vertices before and after the Hadamard product does not effect the result). The second conditions is satisfied just because for a given element of the adjacency matrix to switch its sign it must first cross zero at which point its dynamics must halt indefinitely (if $A(\tau)_{ij}=0$ at some time $T$ then $\frac{d}{d\tau}A(\tau)_{ij} =0$ for all $\tau>T$). This is a direct consequence of a Hadamard product used in \eqref{eq:dimone}. And the third condition is imposed by a the term $ - \hat{A} \hat{A}^{\dagger} \hat{A}$ which protects the eigenvalues of $\hat{A}$ matrix from growing large.  

Now we turn to the fourth condition which is in fact the most difficult to satisfy and also the least transparent. To gain a somewhat better understanding of the graph flow \eqref{eq:dimone} it is convenient to first rewrite it as an equation for eigenvalues   $\lambda_i$   of the matrix $\hat{A}$ with a caveat that these eigenvalues come in conjugate pairs (since the adjacency matrix is real). Strictly speaking the conjugate pairs do not evolve independently of each other because of the Hadamard product in \eqref{eq:dimone}, but for a moment we shall neglect these Hadamard-product-mediated (or simply Hadamard) interaction and rewrite the graph flow equations as a flow of its eigenvalues, i.e.
\be
\frac{d\lambda_i}{d\tau}   \propto \left (\lambda_i - \lambda_i \lambda_i^* \lambda_i  \right ).
\ee
with no summation over repeated indices and $\lambda^*$ is a complex conjugate of $\lambda$. Evidently the above equation describes evolution of particles in a ``Mexican-hat'' potential
\be
V(\lambda) \propto \left (1  - \lambda^* \lambda \right )^2,
\ee
which is the main reason why the form \eqref{eq:dimone} of the graph flow equation was used on the first place. In the long run all of the eigenvalues will attempt to occupy the lowest possible ``energy'' states described by a circle of unit radius on a complex plane 
\be
 \lambda^* \lambda = 1.
\ee
But if we ignore directions of the graph edges and consider the spectrum of an undirected adjacency matrix $\hat{\bf{A}}$ of Eq. \eqref{eq:undirected}, then it would be given by real components of these eigenvalues. And this is exactly what one expects from a one-dimensional periodic lattice. Of course in reality the Hadamard interactions may prevent all of the eigenvalues to occupy the smallest energy state and we should expect some non-trivial arrangements of the degrees of freedom to emerge. 

\begin{figure}
\begin{center}
\includegraphics[width=0.75\textwidth] {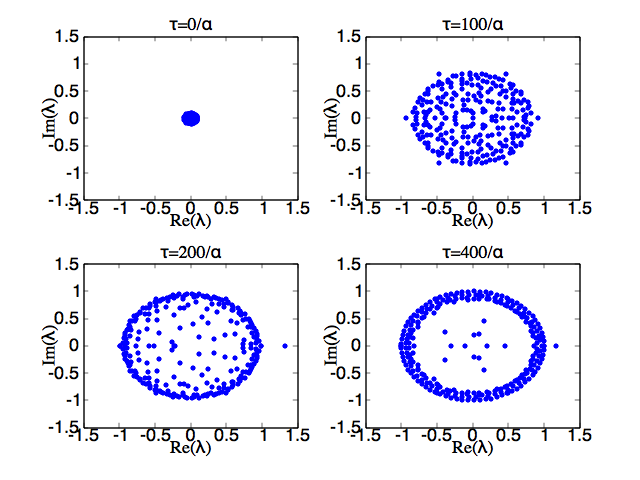}
\caption{Eigenvalues of a $256\times256$ adjacency matrix at four different values of $\tau$ for a graph flow describe by equation \eqref{eq:dimone}.}  \label{plot1}
\end{center}
\end{figure}

To summarize,  what we expect to happen is that the graph flow would attempt to deform the adjacency matrix to that of the one-dimensional lattice but it might not succeed completely because of the non-trivial Hadamard interactions between eigenvalues. On Fig. \ref{plot1} we present results of a numerical simulation of a graph flow with $1D$ geometric attractors. We start with a $256\times256$ adjacency matrix whose elements are chosen randomly (as absolute values of a Gaussian random variable with zero mean and variance $0.01$) and then evolve it using the graph flow equation \eqref{eq:dimone}. Evidently, eigenvalues of the adjacency matrix are attracted towards unit norm and as a result $\hat{A}$ is attracted towards one of the permutation matrices which turned out to be a fixed point of the graph flow, or what we shall call a $1D$ geometric attractor. Each cycle of this permutation matrix describes a one-dimensional periodic lattice (a loop) and deviations of $\hat{A}$ from the permutation matrix are the perturbations whose effective behavior we will discuss in Sec. \ref{Sec:EmergentGeometry}.

\subsection{Colored graphs}\label{coloredgraphs}

We are now ready to tackle a somewhat more difficult two dimensional  problem, but before we proceed let us generalize the graph flow equation \label{eq:GraphFlow} to allow different types (or {\it colors}) of edges between vertices. As we already discussed this requires an introduction of different adjacency matrices that we distinguish by an additional upper (or color) index, i.e. $\hat{A}^{1}, \hat{A}^{2}, \hat{A}^{3}$, etc. With this generalization the graph flow equation(s) takes the following form
\be
\frac{d}{d\tau} \hat{A}^{J} = \hat{f}^{J}(\hat{A}^{J}) + \hat{g}^{J}(\hat{A}^{1},\hat{A}^{2}, ...)\label{eq:GraphFlow2}
\ee
where we have intentionally separated the interaction terms between graphs of different colors. The interaction term might have additional desired symmetries such as invariance under permutations of colors, i.e. invariance of the flow under relabeling of colors of the adjacency matrices. In what follows we consider an example of a graph flow with such a symmetry described by
\bea
&& \frac{d}{d\tau}{\hat{A}}^J = \alpha \hat{A}^J \star \left (\hat{A}^J - \hat{A}^J  \hat{A}^{J\dagger} \hat{A}^J \right ) +  \label{eq:dimtwo}\\
&&+\;  \alpha  \hat{A}^J \star  \sum_{K\neq J} \left ( -\beta \left ( 2 \hat{A}^J -\hat{A}^K \hat{A}^J \hat{A}^{K\dagger} - \hat{A}^{K\dagger} \hat{A}^J \hat{A}^K  \right ) -\gamma\left (\hat{A}^J \hat{A}^K+ \hat{A}^J \hat{A}^{K\dagger} + \hat{A}^K \hat{A}^J + \hat{A}^{K\dagger} \hat{A}^J  \right )\right ) \notag
\eea
 for the simplest non-trivial case of two colors, i.e. $J,K \in \{1,2\}$. The role of the second  (or  $\beta$-term) and third (or $\gamma$-term) terms is to ensure that the system is attracted towards a state for which the smallest non-self-intersecting loop has four edges (due to  $\beta$-term), but there are no loops with three edges ($\gamma$-term). Cleary this is what had to be imposed for the system to flow towards a geometric attractor described (locally) by a square lattice, but for other lattices other interaction terms would have to be used instead. 

\begin{figure}\begin{center}
\includegraphics[width=0.75\textwidth] {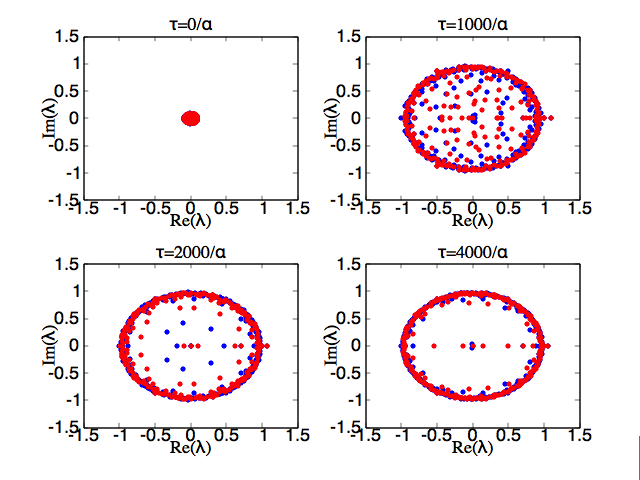}
\caption{Eigenvalues of two $256\times256$ adjacency matrices at four different values of $\tau$ for a graph flow describe by equation \eqref{eq:dimtwo}.}  \label{plot2}
\end{center}
\end{figure}

On Fig. \ref{plot2}
we present  numerical results of a graph flow described by equations \eqref{eq:dimtwo} (with $\beta=1/4$ and $\gamma=1/4$) starting with two adjacency matrices whose eigenvalues are plotted respectively in red and in blue. As the time $\tau$ progresses more and more eigenvalues move closer to a unit circle, although it now takes much longer for the fixed point to be reached. However, even by time $\tau=4000/\alpha$ the graph flow had already taken the adjacency matrices very close to a $2D$ geometric attractor which is evident from Table \ref{Table1}
\begin{table}
 \caption{An average number of closed loops $\mu(n)$ of length $n$ for $1D$ and $2D$ lattices and random graphs after respective graph flows. }
\begin{center}
  \begin{tabular}{ l | c | c | c |  r }
    \hline
     $n$ & 1d lattice & graph flow of Eq. \eqref{eq:dimone}   & 2d lattice &  graph flow of Eq. \eqref{eq:dimtwo} \\ \hline
    1 	& 0 	& 0.01  	& 0 		& 0.01  	  \\ \hline
    2 	& 2 	& 2.00 	& 4 		& 3.99  	 \\ \hline
    3 	& 0 	& 0.03 	& 0 		& 0.20 	 \\ \hline
    4 	& 6 	& 6.00	& 36 		& 35.6  	 \\ \hline
    5 	& 0 	& 0.14 	& 0 		& 3.84 	  \\ \hline
    6 	& 20 	& 20.0 	& 400 	& 394  	 \\ \hline
    7 	& 0 	& 0.60	& 0 		& 65.3  	 \\ \hline
    8 	& 70 	& 70.2 	& 4900 	& 4846  	 \\ \hline
    9 	& 0 	& 2.54 	& 0	 	& 1070  	 \\ \hline
   10 & 252 & 253 	& 63504 	& 63185  	
\label{Table1}
  \end{tabular}
\end{center}
\end{table}
where we compare
\be
\mu(n)\equiv \frac{1}{N}Tr\left (\hat{\bf{A}}^n\right ), \label{eq:moments}
\ee
for different graph flows and lattices. An undirected adjacency matrix of the lattice is now defined as
\be
\hat{\bf{A}}\equiv \sum_J\left ( \hat{A}^J+\hat{A}^{J\,\dagger} \right ).
\ee
The quantity $\mu(n)$ basically counts an average number of closed loops with length $n$ and for small values of $n$ encodes local properties of corresponding lattices.  By comparing the last two columns of the table one can see that after the graph flow of Eq. \eqref{eq:dimtwo} the graph resembles very closely a $2D$ lattice, and, likewise, after the graph flow of Eq. \eqref{eq:dimone} the graph is that of a $1D$ lattice. We have conducted a similar numerical analysis for a $3D$ geometric attractor and can report that essentially the same construction works there as well. 

\section{Emergent Geometry}\label{Sec:EmergentGeometry}

Now that we have verified that certain graph flows are attracted towards (locally) low-dimensional lattices, we can try to better understand the emergence of a low-dimensional geometry. As was already noted, after the graph flow the adjacency matrix $\hat{A}^{J}$ of a given color $J$ has on average one element of order $1$ per row, one element of order $1$ per column and all other elements are much smaller than $1$, i.e. it is approximately a permutation matrix. Of course these permutation matrices (and their cycles) are not uncorrelated as they form together a low-dimensional lattice, and thus we can use these cycles to define a local coordinate system and then rewire the graph flow equation as an approximate field theory equation. 

The (local and discrete) coordinates of a square lattice $\vec{x} = (x^1, x^2, ... )$ (with unit lattice vectors $\vec{e}_{J}$) can be mapped to a subset of vertices $v(\vec{x})$  of the information graph  such that the following condition is satisfied
\be
\vec{y} = \vec{x} +\vec{e}_J  \;\;\;\; \Rightarrow \;\;\;\;\;   A^{J}_{v(\vec{x}), v(\vec{y})} \approx 1.
\ee
Given this map we can also define diagonal elements of a (local inverse) metric as
\be
g^{JJ}(\vec{x}) = A^{J}_{v(\vec{x}), v(\vec{x}+\vec{e}_{J})}
\ee 
and by construction all non-diagonal elements are set to zero. At late times of the graph flow  the metric would be nearly flat and so one can define diagonal perturbations as
\be
 h_{JJ}(\vec{x}) = g_{JJ}(\vec{x}) - 1.
\ee 

\subsection{Geometric flow}

In a one-dimensional case such perturbations are represent by a single function $h(x)$ of a single variable and we can expand Eq. \eqref{eq:dimone} to a linear order in $h$ to obtain 
\bea
\frac{d {h}(x)}{d\tau}  & = & \alpha (1+h(x)+{\cal O}(h(x)^2)) \left ( (1+h(x)+{\cal O}(h(x)^2)) - (1+h+{\cal O}(h(x)^2))^3 \right) \notag \\
& =  & -2 \alpha h(x)+{\cal O}(h(x)^2). 
\eea
This confirms that $h$ must decay exponentially in agreement with our numerical results which showed the $1D$ lattice is an attractor of a graph flow described by \eqref{eq:dimone}. One might however wonder if these perturbations can undergo other evolutions under the graph flow. For example, if we want the perturbations to satisfy a diffusion equation (with escape), then on the lattice it would be given by
\bea
\frac{d{h}(x)}{d\tau} & = & -2 \alpha h(x) + \alpha \delta \left ( h(x) - h(x-1) )  - ( h(x+1) - h(x) )  \right ) \notag\\
& = &  -2 \alpha h(x) + \alpha \delta \left ( - h(x-1) + 2 h(x) -  h(x+1)  \right ). \label{eq:diffusion}
\eea
This can be achieved with a graph flow 
\be
\frac{d}{d\tau} {\hat{A}} = \alpha \hat{A} \star \left (  \hat{A} -  \hat{A} \hat{A}^{\dagger} \hat{A} + \frac{\delta}{2} \left [ \hat{A}, \left [ \hat{A}^{\dagger}, \hat{A}\right] \right ]  \right ) \label{eq:dimone_diff}, 
\ee 
where the commutator is defined as usual 
\be
\left[ \hat{A}, \hat{B} \right] =  \hat{A} \hat{B} - \hat{B} \hat{A}.
\ee

For a $D$-dimensional case diagonal perturbation of the metric are given by $D$ functions $h_{JJ}(\vec{x})$ of $D$ variables $\vec{x}$. Then one might want to define a graph flow so that perturbations obey the Ricci flow equation (with escape) which can be thought of as a generalization of \eqref{eq:diffusion} to higher dimensions. Indeed, in our coordinates the Ricci flow \cite{Ricci} of the metric perturbations to the linear order would be given by uncoupled diffusions equations for each component which takes the following form 
\be
\frac{d}{d\tau} h_{JJ}(\vec{x}) = -(D+1) \alpha h_{JJ}(\vec{x})  + \alpha \delta  \sum_{K} \left (  - h_{JJ}(\vec{x}-\vec{e}_K)+ 2 h_{JJ}(\vec{x}) -  h_{JJ} (\vec{x}+\vec{e}_K) \right )\label{eq:Ricci_flow}
\ee
To obtain this flow of the metric perturbation the graph flow equation  \eqref{eq:dimtwo}  (with  $\beta=\gamma=1/4$) must contain an additional interaction term 
\be
 \alpha \hat{A}^J \star \left (  \frac{\delta}{2} \sum_K \left (\left [ \hat{A}^K, \left [ \hat{A}^{K\dagger}, \hat{A}^J\right] \right ] - \left [ \hat{A}^{K\dagger}, \left [ \hat{A}^{K}, \hat{A}^J\right] \right ]  \right )\right ) \label{eq:dimtwo_diff}.
\ee 

\subsection{Effective dynamics} 

Now that we understand how the metric emerges from the graph flow we can ask what would happen if we slightly change the initial state of the  graph  (e.g. due to a true dynamics of either quantum or statistical system in question). How would that effect the adjacency matrix and thus the emergent metric $g_{IJ}$ after the graph flow? The ultimate goal here is to extract an effective description of the dynamics after the graph flow as an equation for $g_{IJ}(t)$. At this point it might be useful to look back at our chart of Eq. \eqref{eq:Chart} where horizontal direction represents approximation steps (e.g. graph flow with parameter $\tau$) and vertical direction represents evolution with respect to time $t$. 

For starters, we can check how the spectrum of adjacency matrix at final stages of the graph flow evolves (i.e. approximate dynamics on chart \eqref{eq:Chart}) under small displacements of the adjacency matrix in the initial state (i.e. information dynamics on chart \eqref{eq:Chart}). To illustrate the main point the evolution need not be realistic, i.e. we do not need to follow a true unitary dynamics of either quantum or statistical system. Instead, we first generate a random matrix $\hat{A}(0,0)$ whose elements are absolute values of a Gaussian random variable with zero mean and variance $0.01$ (just like what we did in the pervious section), but then we add to it another random matrix $\hat{\Delta}$ multiplied by a time parameter $t$,  i.e.
\be
\hat{A}(t,0) = \hat{A}(0,0) + t \hat{\Delta}. \label{Eq:toy}
\ee
Elements of matrix $\hat{\Delta}$ are also Gaussian variables with zero mean, but much smaller variance $\sigma=0.00002$. Strictly speaking such a `toy' evolution can potentially make some element of the adjacency matrix $\hat{A}(t,0)$ negative, but this is not likely to happen for sufficiently small $\sigma$ and $t$.

\begin{figure}\begin{center}
\includegraphics[width=0.75\textwidth] {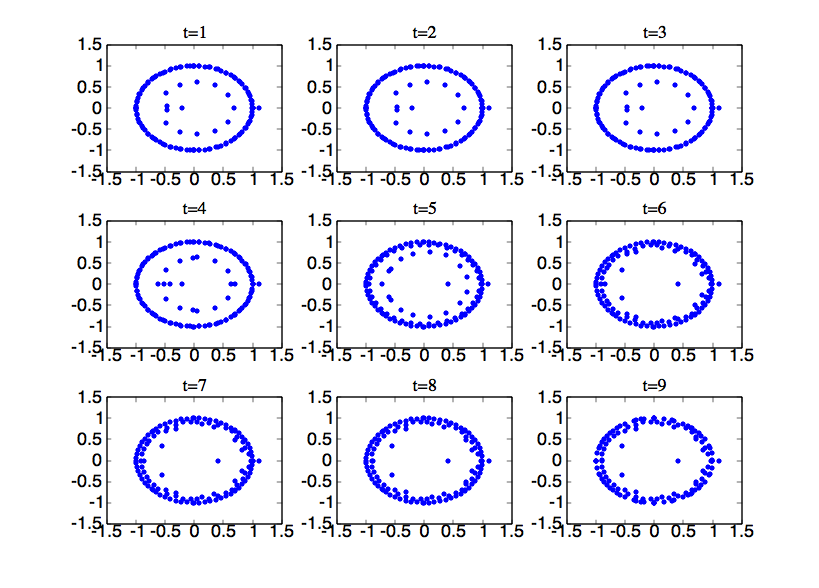}
\caption{Eigenvalues of a $256\times256$ adjacency matrix at a fixed graph flow parameter $\tau=400/\alpha$ and nine different time steps $t=1,...,9$ described by Eq. \eqref{Eq:toy}. Vertical and horizontal axes represent imaginary and real components respectively.}\label{plot3}
\end{center}
\end{figure}

On Fig. \ref{plot3}
we plot eigenvalues of the adjacency matrices at nine different times $t$ all of which were deformed by the graph flow equation \eqref{eq:dimone} until the flow parameter reaches $\tau=400/\alpha$. Note that from $t=1$ to $t=4$ the spectrum is not significantly modified, from $t=4$ to $t=6$ there is a sudden jump in the spectrum, and from $t=6$ to $t=9$, once again, the spectrum is only slightly perturbed. What must have happened is that by changing the initial state from $\hat{A}(4,0)$ to $\hat{A}(6,0)$ the system switched from one $1D$ geometric attractor to another  $1D$ geometric attractor. It is important to note that although the spectra of the first four (and also the last four) plots are similar the adjacency matrices could be very different  (although they should be approximately related to each other through permutation of vertices transformations).  What about an emergent dynamics of the metric perturbations $h^{IJ}(t)$? We have not analyzed it here, but it should not be too difficult at least in the regime where the graph flow does not switch between geometric attractors. 

\section{Conclusion}\label{Sec:Conclusion}

In this paper we described a procedure which enables us to approximate either quantum or statistical systems as geometric objects. The geometric approximation procedure consisted of three steps: \\ (1) construction of information graphs from a quantum or statistical system (see Sec. \ref{Sec:Information Graph}),\\ (2) evolution of information graphs using the graph flow equations (see Sec. \ref{Sec:GraphFlow}), and\\ (3) derivation of an approximate geometric description after the graph flow (see Sec. \ref{Sec:EmergentGeometry}). \\At the end of the process one obtains a discrete lattice with an approximately flat geometry on the smallest scales, but a (generically) non-trivial geometry and topology on larger scales.

This suggests a possible approach to (a non-perturbative) quantum gravity in which the geometry (a secondary object) emerges directly from a quantum state (a primary object) due to the geometric approximation procedure. Up to this point we have only discussed the emergence of a metric with Euclidean signature, but for quantum gravity one would need a lot more. In particular we need to understand how  a metric with Lorentzian signature might emerge. One idea is that an effective dynamics of the metric would be due to a unitary evolution of a quantum state (for an illustration of this point see the chart \eqref{eq:Chart}). Of course the metric is only one type of local degrees of freedom that might emerge near fixed points of a particular graph flow, but there could also be other (we call them matter) degrees of freedom whose evolution would be governed by other graph flow equations. Then the main question should be how these degrees of freedom evolve in time $t$. In particular is there  a unitary evolution of a quantum state $|\psi\rangle(t)$ and a graph flow equation which gives rise to an approximate evolution of the (metric and matter) degrees of freedom described by the Einstein field equations? And if so can we identify a principle that would single out a unique $\hat{f}$, or more generally $\hat{f}^J$'s and $\hat{g}^J$'s, in the graph flow equations \eqref{eq:GraphFlow} and \eqref{eq:GraphFlow2}? A set of necessary conditions that a graph flow must satisfy was already listed in Sec. \ref{Sec:GraphFlow}, but we can probably do better than that. Perhaps what one also wants to demand is that the least amount of information is lost during this smoothing process. Without going into details we can call it the principle of {\it minimum information loss}. For example, one might demand that the graphs before and after flow are as close to each other as possible according to some metric of distances between graphs, e.g. cut-distance  \cite{Cut} or spectral-distance \cite{Spectral}. Then it would be interesting to see which metric of distances is preferred for the emergence of general relativity and of the observed universe from the minimum information loss principle.  

Another idea is to forget about temporal evolution $t$ and to define a graph flow of four adjacency matrices: three with only non-negative elements (representing three spacial directions) and one with only non-positive entries (representing a temporal direction). This should break the color relabeling symmetry introduced in Sec. \ref{Sec:GraphFlow}, but the hope is that some kind of a Lorentzian symmetry would emerge near fixed points of the graph flow. These four adjacency matrices would represent geometry and there could still be other adjacency matrices which would represent matter. In the previous section we showed that near fixed points the metric dynamics could be often approximated by some geometric flow (e.g. Ricci flow), and we expect (speculate) that dynamics of there matter degrees of freedom could potentially be approximated as a renormalization group flow of some effective field theories. But then if we rewind the graph flows parameter $\tau$ to smaller values the two flows (i.e. geometric flow and renormalization group flow) would start interacting. At this point we generically expect that both metric and matter degrees of freedom would be coupled, but one could still hope that there would be a valid effective (low-dimensional) description of this more general flow. And if so, once again, could it be the Einstein field equations? 

And finally, we note that the fundamental building blocks of Hilbert space in loop quantum gravity \cite{Rovelli} --- spin networks \cite{Penrose} --- are also colored/directed graphs and so it is tempting to apply the graph flow procedure developed in this paper to spin networks. A generic quantum state in loop quantum gravity is a superposition of all possible spin networks and so the same graph flow would have to be applied to each spin network separately. As a result, the original state would be deformed to (or approximated by) a state in a (tiny) subspace of the Hilbert space spanned by only spin networks which represent low dimensional lattices, i.e. semi-classical geometries. This, however, leaves a lot of room for superpositions between semi-classical geometries and, when the time is switched on, to transitions between different geometries. It would also be interesting to see if the graph flow may be responsible for the emergence of general covariance in the models of loop quantum gravity. 

In conclusion, we would like to emphasize that the graph flows can also be used for the purpose of graph sparsification which is of a key importance in machine learning and network science. For example, starting with a collection of weighted directed graphs (which need not be information graphs) one might wish to make them flow towards low-dimensional lattices using the graph flow equations introduced in the paper (we called it a geometric sparsification). Of course, there are (infinitely) many other graph flow equations which do not have geometric attractors,\footnote{The graph flows which satisfy conditions (1), (2) and (3) from Sec. \ref{Sec:GraphFlow}, but not conditions (4) and (5)} but nevertheless could lead to very sparse adjacency matrices and so it would be interesting to figure out which graph flows preserve either cut similarity  \cite{Cut} or spectral similarity \cite{Spectral} of graphs.

{\it Acknowledgments.} The author is grateful to Daniel Harlow, Mudit Jain, Mahdiyar Noorbala  and Evan Severson for useful discussions and comments on the manuscript. The work was supported in part by Templeton Foundation and Foundational Questions Institute (FQXi).

\end{document}